\documentclass[a4paper]{article}

\usepackage[english]{babel}
\usepackage{amsmath}
\usepackage{graphicx}
\usepackage{fullpage}
\usepackage{enumerate}

\title{SIMPLE  EXPLANATION  ON  WHY  QKD  KEYS  HAVE  NOT  BEEN  PROVED SECURE}

\author{Horace P. Yuen\\Department of Electrical Engineering and Computer Science\\Department of Physics and Astronomy\\Northwestern University, Evanston Il. 60208\\email: yuen@eecs.northwestern.edu}

\date{}

\begin{document}
\linespread{1}
\maketitle

\begin{abstract}
A simple counter-example is given on the prevalent interpretation of the trace distance criterion as failure probability in quantum key distribution protocols. A summary of its ramifications is listed.
\end{abstract}

\vspace{2mm}

Quantum key distribution (QKD) is widely perceived to have been proved ``secure" in various protocols, in contrast to conventional encryption methods. In particular, perfect security is taken to hold with a high probability [1] for finite concrete protocols. In this short note we give a simple explanation on why that is not the case and summarize some of the major ramifications. The reader can trace some details from the literature.

Protection of data privacy has become an increasingly important problem that affects even our daily lives.
Let Adam and Babe be two users with private communications and Eve an attacker who wants to learn about the communication content. Protection can be obtained by the mod-2 addition of each ``plaintext" data bit $X_i$ by a key bit $K_i$ to form the ``ciphertext" bit $Y_i$, $i$ the index for the sequence. If the bits $K_i$ are uniformly distributed, i.e., each $K_i$ takes the bit value 0 or 1 with probability 1/2 and is statistically independent of other $K_i$ bits, perfect secrecy on $X_i$ is obtained in that Eve who intercepts the ciphertext cannot tell what $X_i$ is, other than the guess it is 0 or 1 with probability 1/2. This ``one-time pad" encryption requires pre-shared sequence $K=\{K_i\}$ between the users to be as long as the data sequence $X=\{X_i\}$, which is impractical in most application situations. In conventional cryptography a cipher with short shared secret key $K$ that is often obtained by public-key method, is used as the seed key of an encryption cypher in lieu of one-time pad. The security of such approach depends on the unproved complexity based security on the public-key generation of $K$ and the unproved security of the encryption cipher with a short seed key. QKD proposes to provide ``information theoretic security" for the generation of long $K$ which is perfect (the uniform random variable $U$ to Eve) and can be used in ``one-time pad" encryption of the data $X$.

However, it is not possible to guarantee the generation of a perfect key from QKD. The use of a quantum trace distance criterion $d$ is currently adopted [1] to measure the deviation from a perfect key $K=U$. The interpretation [2] is that $d$ gives the maximum ``failure probability" of the QKD scheme, and ``failure" means the key is not perfect. Thus under $d$ the QKD generated key $K$ is asserted to be the random $U$ to Eve except for a probability $d$. The best value of $d$ in theory that has been obtained thus far [3] for a vanishing key generation rate is $\sim 10^{-14}$, and experimentally for $\sim 100$ Kbps key rate it is $\sim 10^{-9}$ [4].  

It turns out that in the case of ``failure" the key may be totally compromised [5] in that $d$ is essentially the probability that Eve may estimate the whole $K$ correctly and hence recover the entire plaintext sequence $X$ from the ciphertext sequence $Y$, the latter always taken to be known to an attacker. Assuming the failure probability interpretation of $d$ is correct, which we'll show is not, a level of even $d\sim 10^{-14}$ may be far from adequately secure depending on the total number of rounds carried out. The derived $d$ level is actually an average over the family of privacy amplification codes. In terms of individual probability guarantee the effective $d$ level becomes $d^{1/2}$. According to the correct interpretation, it becomes $d^{1/3}$ for ciphertext-only attacks and $d^{1/4}$ for known-plaintext attacks [6]. Thus, the above theoretical and experimental values are not adequate for meaningful security guarantee.

Next, we will show $d$ is not the probability of $K$ being not perfect. Generally $K$ is not perfect with probability 1 [6,7] and not $d$ [1]. Consider \textit{known-plaintext attack} on $K$ when $K$ is used as ``one-time pad" encryption, which can be readily launched in at least commercial communications. In such a scenario, Eve knows a portion of the data, say the first $m$ bits of $\{X_i\}$. From the corresponding ciphertext $\{Y_i\}$, she would get the first $m$ bits of $\{K_i\}$. Since $K$ is not perfect, there may be correlations between these first $m$ bits of $\{K_i\}$ and the rest of the $K_i$'s. Thus Eve would learn something about the rest of $\{X_i\}$ from the $\{Y_i\}$ and such $K_i$ correlation.

It is easy to see how a key $K$ with security level $d$ may be totally compromised in a known-plaintext attack. The trace distance reduces to a statistical distance guarantee $\delta_E$ between Eve's probability distribution $\{P_i\}$ on $K$ and the uniform distribution $U$
\begin{equation}
\delta_E=\frac{1}{2}\:\Sigma_i\:|P_i-U_i|
\end{equation}
From her attack Eve obtains her probability distribution $\{P_i\}$ on the $N=2^n$ possible values of the $n$-bit QKD generated key $K$ [5]. The security level from $d$ guarantees only that Eve's $\{P_i\}$ must satisfy (1) with $\delta_E$ no bigger than $d$.

Consider the case where a specific first $m$ bits in $K$ determines the rest uniquely so that Eve's $\{P_i\}$ is transformed from $U$ as follows. Let the specific first $m$ bits with one unique sequence of following bits which  absorbs the total probability of all the $2^{n-m}$ following bits, and let the other bit sequence remains at the $U$ level of $2^{-n}$. Then from (1) it is easily calculated that $\delta_E$ for such a distribution $\{P_i\}$ is $2^{-m}$. Thus, at $d=2^{-m}$, knowing $m$ bits in a known-plaintext attack may totally compromise the key $K$ with probability 1. On the other hand, from the ``failure probability" interpretation of $d$ with or without the ``universal composition" claim of [2], the failure probability remains at $d$ regardless of what $m$-bit portion of $K$ is known to Eve. This counter-example shows explicitly the failure probability interpretation of $d$ is incorrect. At $d\sim 10^{-14}\sim 2^{-46}$, it is not guaranteed that the whole key $K$ is not compromised from knowing just 46 bits of data. Generally, there are many possible partial leaks with possible high probabilities for given known bits of $K$ from a known-plaintext attack which need to be protected against.

We would not repeat our discussions [5,6,7] on how the incorrect interpretation was invalidly derived and
what the general situation of Eve's distribution $\{P_i\}$ actually is under a $d$-level guarantee. A stronger condition has to be imposed on the $\{P_i\}$ for it to hold which is not covered by a fixed $d$ level [6]. However, it may be mentioned that another reason was offered in support of the wrong $d$ interpretation through ``indistinguishability" between the actual and the ideal cryptographic situations, which confuses the mathematical meaning of such ``indistinguishability" and its daily life meaning (plus the Leibniz principle on the identity of indiscernibles). Sometimes security is taken to be a matter of definition or stipulation through terms like ``$\epsilon$-secure", although operational security meaning must be provided for any criterion.

While the wrong interpretation gives too strong a guarantee, for a given $d$ level one can prove an average guarantee on Eve's success probabilities over the known $m$ bits of $K$ [6]. In the above counter-example there is a probability guarantee through the average guarantee, at an effective $d \sim 10^{-14/4} > 10^{-4}$. So the $d$ criterion may be fine if its guaranteed level can be made much smaller than that currently predicted.

In QKD security proofs without known-plaintext attack, already the $d$ level derived is a double average, averaged over the possible privacy amplification codes and over the possible measurement results of Eve from her probing attack. When Markov inequality is used to convert an average guarantee to a probabilistic guarantee, the effective probability guarantee is changed to $d^{1/3}$. With the additional average over the known portion of $K$ in a known-plaintext attack, it becomes $d^{1/4}$ [6]. 

Why is probability guarantee needed and not just average? Because probability is a more operationally meaningful guarantee, as probability and not average is used in the quality control of common product manufacturing. Say, consider the situation where with 50/50 chance a cryptosystem is either secure for 100 years or not secure at all, with an average of 50 years security. Furthermore, too little rounds would be made in comparison to the large exponential number of possibilities in the case of privacy amplification or measurement result averages in ciphertext-only attacks, and possibly in known key bits average also in known-plaintext attacks. With such relatively small number of trials, average has no operational empirical meaning because there are far from enough trials for each possibility to appear just once.

Thus, in operational probability terms the effective $d$ guarantee for cipher-text only attack is $d>10^{-5}$ at best theoretically and $d\sim 10^{-3}$ experimentally. For known-plaintext attack it becomes $d > 10^{-4}$ and $d>10^{-2.5}$. Even at a much lower $d$ level there is security problem when such key is used in an application as in message authentication or for the next QKD round [6]. It appears impossible to extend these numbers by the many tens of order of magnitude either in theory or in practice in order to provide meaningful security guarantee. If a probability of $10^{-15}$ for a single trial is to be considered practically impossible and there are just $10^{5}$ rounds, $d$ would need to be $10^{-80}$ for the effective $d^{1/4}$ level to be $10^{-20}$ for security guarantee. For ciphertext only attacks, obtainable QKD security level is very unfavorable compared to the information theoretic security that can be obtained in symmetric key expansion [7]. It seems quite inappropriate to call the guarantee that can be obtained in QKD as ``unconditional security", particularly because there is no ``security parameter" in QKD with which the security can be pushed to become arbitrarily close to perfect at any fixed key rate. Eve's mutual information on $K$ is itself not a good quantitative criterion [1,6]. Furthermore, that it goes to zero asymptotically in the bit length $|K|$ does not imply $|K|$ is a security parameter [5,6].

Quantification of security with error correction is an almost insurmountable problem. For a given error correction procedure it is not possible to rigorously estimate how many errors Eve could correct for herself, and with what probability, from knowledge of related public announcement. No justification has been given for the common $leak_{EC}$ formula that is widely employed [6]. Even if the parity check digits of a linear error correcting code are covered by true one-time pad, there is still the leak of the code structural information. More significantly, with a necessarily imperfect key used to cover such digits the resulting security is uncertain, especially at the relatively large $d$ levels that can be obtained. The seriousness of the problem can be seen from the use of imperfect keys in message authentication, in which the tag length security parameter is lost due to the imperfect key [6].

We summarize the main consequences of an imperfect key from the incorrect interpretation of the trace distance criterion:
\begin{enumerate}[(i)]

\item The effective trace distance level for operational individual guarantee is reduced from $d^{1/2}$ to $d^{1/3}$  for ciphertext-only attack, and to $d^{1/4}$ for known-plaintext attack when the key is used for encryption.

\item With an imperfect key there is no valid quantification on the effect of error correction, in particular on any $leak_{EC}$ subtraction in a security proof.

\item When the key is not perfect there are other problems such as the number of bits Eve may get correctly even though she gets a sequence incorrectly. Such problems could lead to some not yet discovered ``distinguishing attack" for a nonideal key [8].

\item The imperfect key leak may accumulate in an application similar to the case of message authentication.

\end{enumerate}

The foundational consideration in this paper applies to all QKD protocols, including the recent measurement-device-independent approach, and it shows that no unconditional security can yet be validly asserted. On the other hand, proof of security is the widespread QKD claim that distinguishes it from other known cryptographic methods. It appears that some new ingredient needs to be introduced for provable security. This note should be ended with the following words from [8, p.19], ``soon realize firsthand that security is incredibly subtle and that it is very easy to overlook critical weaknesses."

\begin{center}
\textbf{Acknowlegement}
\end{center}
I would like to thank O. Hirota, G. Kanter, and L. Kish for their useful suggestions on this manuscript.

\end{document}